\input harvmac.tex
\noblackbox
\input epsf.tex
\overfullrule=0mm
\newcount\figno
\figno=0
\def\fig#1#2#3{
\par\begingroup\parindent=0pt\leftskip=1cm\rightskip=1cm\parindent=0pt

\baselineskip=11pt
\global\advance\figno by 1
\midinsert
\epsfxsize=#3
\centerline{\epsfbox{#2}}
{\bf Fig. \the\figno:} #1\par
\endinsert\endgroup\par
}
\def\figlabel#1{\xdef#1{\the\figno}}
\def\encadremath#1{\vbox{\hrule\hbox{\vrule\kern8pt\vbox{\kern8pt
\hbox{$\displaystyle #1$}\kern8pt}
\kern8pt\vrule}\hrule}}

\Title{USC-96-009}
{\vbox{
\centerline{Correlations in one dimensional quantum impurity
}
 \vskip 4pt
\centerline{problems with an external field.}
}}

\centerline{F. Lesage, H. Saleur.}
\bigskip\centerline{Department of Physics,}
\centerline{University of Southern California,}
\centerline{Los Angeles, CA 90089-0484}
\centerline{PACS: 71.10.Pm, 71.15.Cr, 72.10.Fk.}

\vskip .3in

We study  response functions of integrable quantum impurity problems
with an external field at $T=0$ using non perturbative
techniques
derived from the Bethe ansatz. We develop the first steps
of the theory of excitations over the new, field dependent ground
state, leading
to renormalized (or ``dressed'') form-factors. We obtain exactly the
low
frequency behaviour of the dynamical susceptibility  $\chi''(\omega)$
in the double well problem of
dissipative
quantum mechanics (or equivalently the anisotropic Kondo problem),
and the low frequency behaviour of the AC noise
$S_t(\omega)$ for tunneling between edges
in  fractional quantum Hall devices.
We also
obtain exactly the structure of singularities in $\chi''(\omega)$ and
$S_t(\omega)$.
Our results differ significantly from previous perturbative
approaches.

\Date{10/96}

\newsec{Introduction.}

One dimensional quantum impurity problems, which  are of considerable
physical importance,  have  recently been the subject of significant
theoretical progresses based on their integrability. In particular,
in a recent paper   \ref\LSS{F. Lesage, H. Saleur, S. Skorik, Phys.
Rev. Lett. 76 (1996) 3388, cond-mat/9512087 ;  Nucl.
Phys. B474
(1996) 602,
cond-mat/9603043 .}, we have shown
how correlation functions at $T=0$ and without an applied field
could be obtained
\foot{More precisely, with controlled accuracy all the
way from small to large coupling.} using  massless  form-factors
 \ref\smirnov{F.A. Smirnov, ``Form factors in completely integrable
models of quantum field theory", Worlk Scientific, and references
therein.}
\ref\giu{G. Delfino, G. Mussardo, P. Simonetti, Phys. Rev.
D51, 6620 (1995).}.  As a result, the response function
$\chi''(\omega)$
in the double well problem of dissipative quantum mechanics
\ref\dqmrev{A.J. Leggett,
S. Chakravarty, A.T. Dorsey, M.P.A. Fisher, A. Garg,
W. Zwerger, Rev. Mod. Phys., 59, 1 (1987).} \ref\guinea{
F. Guinea, V. Hakim, A. Muramatsu, Phys. Rev. B32,
4410 (1985); S.A. Bulgadaev, JEPT vol. 38, 264 (1984),
Sov. Phys. JETP, vol. 39, 314 (1984).}, 
and the frequency dependent conductance
$G(\omega)$ for tunneling between edges in  $\nu={1\over 3}$
fractional quantum Hall devices \ref\hallpt{K. Moon, H. Yi, C.L.
Kane,
S.M. Girvin, M.P.A. Fisher, Phys. Rev. Lett. 71, No26, 4381.}
\ref\flsbig{P. Fendley, A.W.W. Ludwig, H. Saleur, Phys. Rev.
B52, 8934 (1995), cond-mat/9503172.}
were obtained at $T=0$ and without
an applied field.
Quantities at finite temperature and with an applied field are also
of great interest, and potentially closer to experiments, but
significantly more difficult to obtain, except for their
DC component \ref\flsnoise{P. Fendley, A.W.W. Ludwig, H. Saleur,
Phys. Rev. Lett. 75, 2196 (1995),  cond-mat/9505031}, \ref\flnoise{P.
Fendley,
H. Saleur,
to appear in
Phys. Rev. B., cond-mat/9601117.}.

In this paper, we address the
case $T=0$ but the applied field non-zero (this
field we designate generically by $V$). We consider in particular
$\chi''(\omega)$ in the double well problem with a bias (equivalently
the anisotropic Kondo problem with a field applied to the
impurity), and
the non-equilibrium noise $S_t(\omega)$ for the tunneling current
in the quantum Hall effect
in the presence of an applied voltage \ref\CFW{C. de C. Chamon, D.
Freed, and X.G. Wen,
Phys. Rev. B51 (1995) 2363.}\ref\CFWI{C. de C. Chamon, D. Freed, and
X.G. Wen, Phys. Rev. B53 (1996) 4033.}.

The problem (and the origin of physically interesting
phenomena) is that the applied  field changes the structure
of the ground state. The new ground state (we refer to it
as the Fermi sea in what follows)  can be obtained
without much difficulty from the   massless basis
 of solitons, antisolitons
and breathers with factorized scattering. It is simply made
of right moving (R) solitons and left moving (L) antisolitons filling
all rapidities up to a Fermi value $A$.
Excitations
are obtained by adding particles and making holes in the sea.
Due to the interacting nature of the theory, the scattering of
these excitations will be different
from the case $V=0$: there is a ``dressing'' of the S-matrices
coming from the Fermi sea, a
 phenomenon largely similar to the dressing observed in lattice
regularizations \ref\korbook{V.E. Korepin,
N.M. Bogoliubov, A.G. Izergin, ``Quantum inverse scattering
method and correlation functions", Cambridge University Press,
(1993).}. To develop a  form-factor approach for the
computation of correlations, we have then to compute dressed
form-factors: the matrix
elements of physical operators between asymptotic states of the
dressed excitations. This is a difficult problem,
for technical reasons which will
be explained later.  Exact results can still be obtained
however. Here, we study in details two questions which have
been the subject of interest
in the litterature:
the low frequency behaviour of the response function in
dissipative quantum mechanics and  the frequency
dependent shot noise in the tunnelling Hall  problem
 \CFW,\CFWI. In addition,  we will explore the
possible existence of singularities at finite frequency
in these quantities.

\fig{Multi-layered Fermi sea, only antisolitons are
filled at T=0.}{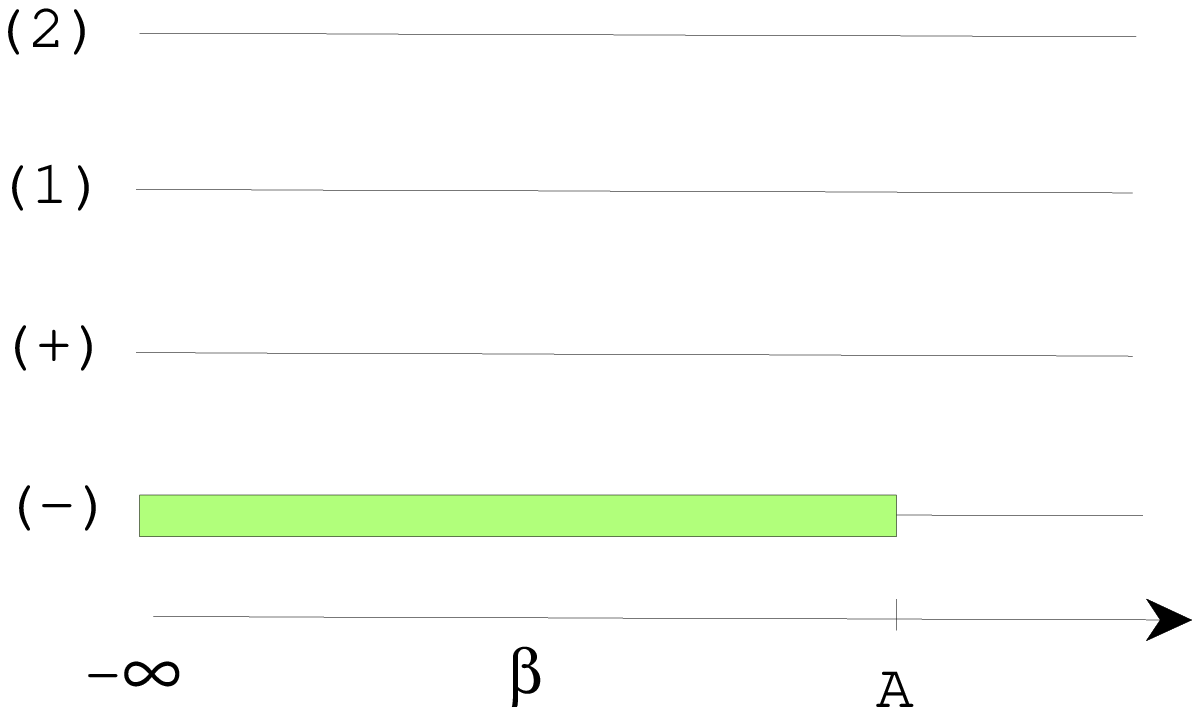}{7cm}
\figlabel\tabb

The hamiltonian for the models we study here is~:
\eqn\hamil{
H={1\over 2}\int_{-\infty}^0 dx \ [8\pi g\Pi^2+{1\over 8\pi g}
(\partial_x\phi)^2]+H_B.
}
This is the standard Luttinger liquid bosonic hamiltonian with
a term at the boundary $x=0$\ref\fisher{C.L. Kane, M.P.A. Fisher,
Phys. Rev. B46, No23, 15233.}.
Both the impurity tunneling   and the double well problem  can be
written into this form by standard manipulations.  The boundary
term depends on the problem, for the Hall effect it is~:
\eqn\hallbound{
H_B=\lambda \left[e^{-i gVt}e^{i\phi(0)/2}+e^{i
gVt}e^{-i\phi(0)/2}\right],
}
and for dissipative quantum mechanics we have~:
\eqn\dissipbound{
H_B=\lambda \left[ S_+ e^{i\phi(0)/2}+S_-
e^{-i\phi(0)/2}\right]+{\epsilon\over 2} S_z.
}
In \hallbound\ $V$ is the voltage. The electron charge is set to
$e=1$,
and the phases in the boundary (tunneling) term correspond to
tunneling of Laughlin quasi-particles. In \dissipbound, $\epsilon$ is
the
bias, and in this formulation, $g=1,1/2$ correspond
respectively to the isotropic and Toulouse points. Let us
recall that \dissipbound\ describes also the anisotropic Kondo
problem
with a field applied to the impurity.

It is sometimes more appealing physically to ``unfold'' \hallbound.
Indeed,
writing the field $\phi$ as a sum of a left and a right
moving part, the integral  on $[-\infty,0]$ can be traded
for an integral on $[-\infty,\infty]$, the hamiltonian acting only
on right movers. The boundary term then becomes an impurity term
with the replacement $\phi(0)\to 2\phi_R(0)$.

Let us now define the
quantities we will study in this language.
For the impurity tunneling, we want to study the non-equilibrium
quantum  noise.
In the  four terminal geometry (see fig. 2),
this noise has several components. For simplicity,
we will discuss in this introduction the tunneling noise only.
Calling $I_t(t)$ the tunneling current, we consider~:
\eqn\shotcorr{
S_t(\omega)=\int_{-\infty}^\infty dt \ e^{i\omega t}
<\{I_t(t),I_t(0)\}>.
}
With respect to figure 2, the tunnelling current is defined
as $I_t(t)=j_R(0^+,t)-j_L(0^-,t)$.

\fig{Four terminal system in the tunnelling $\nu=1/3$ Hall effect.}
{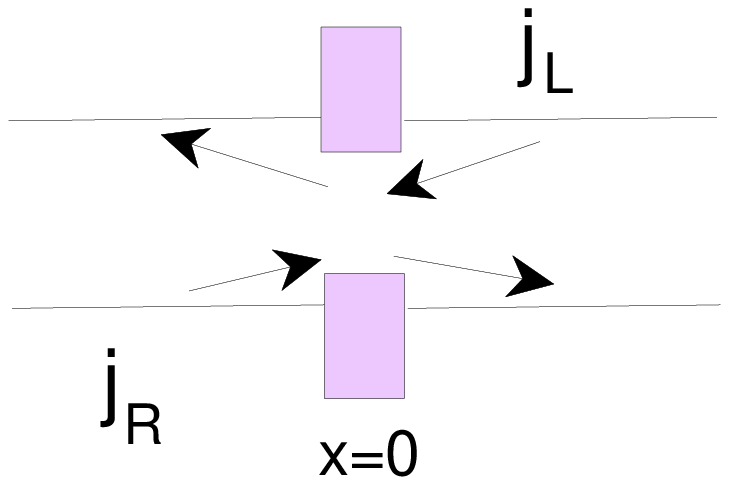}{6cm}
\figlabel\tabb

For dissipative quantum mechanics (or the
anisotropic Kondo problem), the quantity we want to
compute is the dynamical succeptibility\foot{
We use conventions in which $\hbar=1$ and the spins are normalised
to 1.}~:
\eqn\dissipcorr{
C(t)={1\over 2}<[S_z(t),S_z(0)]>
}
or its Fourier transform $\chi''(\omega)$ \foot{Our normalization
is $\chi''(\omega)=\int {dt\over 2\pi} e^{i\omega t} C(t)$ .} \dqmrev
{}.

As described in \LSS\ the hamiltonian with the boundary interaction
\hallbound\
is better understood in the framework of the
sine-Gordon theory.  The idea is simple:  if we add a term
$\Lambda \cos\phi(x)$ to the hamiltonian, then the resulting
problem is known to be integrable, this is the  boundary sine-Gordon
model \ref\ghozamo{S. Ghoshal, A. Zamolodchikov,
Int. J. Mod. Phys. A9, 3841 (1994).}.
This model is conveniently addressed in the basis of
solitons/antisolitons,
breathers, where the bulk  interaction reduces
to factorized scattering encoded in an  $S$ matrix. The boundary
interaction in this basis
is then  described by a  reflection matrix.  In order
to describe the problems we want to study, we then let $\Lambda\to 0$
, to get a ``massless scattering'' description of a free boson,
which involves
massless particles  still referred to as  solitons/antisolitons and
breathers. Again, the
boundary is simply described by a reflection matrix. A similar
description works for \dissipbound.

Having this basis, it is  well known how to compute
the ground state particle densities by using
the Bethe ansatz. Knowing this ground state, a naive
approach to correlations would be to consider
matrix elements \foot{Here, as in our previous works,
we normalize asymptotic states such that
$<\theta|\theta'>=2\pi\delta(\theta-\theta')$.} of the form~:
\eqn\mauvais{
{}^{\gamma_1,...,\gamma_q}
<\alpha_1,...,\alpha_p| {\cal
O}|\beta_1,...,\beta_q>_{\epsilon_1,...,\epsilon_q}
}
where bra and ket stand for the (shifted) ground state plus some
excitations, and to
compute \mauvais\ using crossing. There are
however important difficulties
in this approach which have to do with the
existence of interactions (ie a non trivial
S-matrix).  A more natural, and physically
appealing, way to
proceed is to think in terms of dressed or renormalised excitations
{\it over}
the  ground state.

The idea again is simple, though technically still difficult:  with
the help of
the Bethe equations it is possible to compute the
dressed  energy  and dressed scattering matrix between
excitations.  Having these quantities, one can then try to
write axioms for the dressed form-factors, in analogy with the
well known case when $V=0$ \smirnov, and then solve these axioms
 (a crucial difference
is that in the present case, there is an energy scale $V$, and
relativistic invariance is broken).
Correlators will then follow by inserting a complete set of
excitations
between operators.

The paper is organized as follows~:
In the second section, we study the new ground state
and the
structure of excitations of the auxiliary hamiltonian~:
\eqn\hamref{H={1\over 2}\int_{-\infty}^\infty dx \ [8\pi g\Pi^2+
{1\over 8\pi g}(\partial_x\phi)^2]
+{V\over 4\pi} \int_{-\infty}^\infty \partial_t\phi.}
This hamiltonian is related with the problems of interest through
simple manipulations
described in the appendix, with the correspondence $\epsilon=gV$ for
the dissipative quantum mechanics case.

In the third section we discuss in considerable details the physics
at low energies. We
show that it is described by an effective free-fermion theory with
renormalized parameters. We also discuss the physics near the
thresholds of
various excitation processes, laying the ground for further
discussion of the singularities.

The low frequency behaviour
being completely under control, we are then
able to obtain some exact results as $\omega\to 0$. In the fourth
section
we discuss the behaviour of the
dynamical succeptibility in the double well problem (or
the anisotropic Kondo problem), and we obtain
 a closed expression for $\lim_{\omega\rightarrow 0}
 {\chi''(\omega)\over \omega}$ as a function of $V$ (recall
$\epsilon=gV$,
$\epsilon$ the bias). We prove that Shiba's relation
\ref\shiba{H. Shiba, Prog. Theor. Phys. 54, (1975) 967.}
\ref\shibaweiss{M. Sassetti, U. Weiss, Phys. Rev. Lett. 65,
2262 (1990).}~:
\eqn\shiint{
\lim_{\omega\rightarrow 0}{\chi''(\omega)\over \omega}=\pi^2 g
\chi_0^2
}
holds for system with bias also\foot{Note that this differs
from the usual relation $\chi''/\omega=2\pi g \chi_0^2$ because
of the normalisations of the spins $S_z$ to one and
a factor $2\pi$ in the definition of the Fourier transform .}. A
closed
expression of $\chi_0$ follows, which can be expanded as a series in
${\lambda^2\over V^{2(1-g)}}$ at large voltage (small coupling
$\lambda$),
and in powers of  ${V\over \lambda^{1/1-g}}$ at   small
voltage (large coupling).
There is in particular  a universal product involving small and large
voltage properties
of the static succeptibility $\chi_0$~:
\eqn\univgintro{
\left[\lim_{V\rightarrow 0} \chi_0(V)\right]^{2(1-g)} \times
\left[\lim_{V\rightarrow\infty}
V^{3-2g} \chi_0(V)\right]=F(g)
}
where the function $F$ is given
explicitely in the text.

These behaviours can be verified numerically
using the
framework of the Numerical Renormalisation Group method
(NRG) \ref\costi{T.A. Costi, C. Kieffer, Phys.
Rev. Lett. 76, 1683 (1996).} which is very precise,
and indeed the large field exponent
found there
is in agreement \ref\costipriv{T.A. Costi, Private Communication.}
with the previous result.
They could possibly also be checked using monte carlo
simulations \ref\chakra{S. Chakravarty, J. Rudnick,
Phys. Rev. Lett. 75 501 (1995).}, and maybe experimentally.
Early results on $\chi''(\omega)$, without bias, were also found in
\ref\guineanum{F. Guinea, Phys. Rev. B, vol. 32,
4486 (1985).} using sum rules and scaling arguments.

In the fifth section we discuss the noise in the tunneling
problem. We obtain in particular  the results~:
\eqn\rnoise{\eqalign{
\lim_{\omega\to 0}{S_t(\omega)\over \omega}&= {g\over \pi} , \
V<<T_B\cr
\lim_{\omega\to 0}{S_t(\omega)\over \omega}&=K {g\over \pi}V^4
 \left({\lambda\over V}\right)
^{4/g}, \ V>>T_B .
}}
Expressions for the various components
of the noise in the four terminal geometry as well as the noise
in the total current  are also obtained.

In sections 4  and 5 we also discuss the existence of potential
singularities, restricting to $1/g$
an integer strictly larger than 2. We
find that $\chi''(\omega)$ as well as $S_t(\omega)$  should have
singularities at all values $\omega=ngV$, $n$ an integer. We show
that
the first singularity is a discontinuity
in the first order derivative, ie  of the form $|\omega-gV|$  (recall
$\epsilon=gV$).
We argue that
the other singularities should  be of the same form: $|\omega-ngV|$,
although
we cannot completely prove it.
All of these singularities disappear when
$g=1/2$ where $\chi''$ is regular, while $S_t(\omega)$ has
a weaker residual singularity at $\omega=V$, of the form
$|\omega-V|^3$.

\newsec{The Fermi sea and the structure of excitations at T=0.}

For both problems, some simple manipulations (see the appendix) lead
first to the consideration of the bulk hamiltonian~:
\eqn\hbulk{H={1\over 2}\int_{-\infty}^\infty dx \ [8\pi
g\Pi^2+{1\over 8\pi g}
(\partial_x\phi)^2]+{V\over 4\pi}\int_{-\infty}^\infty
\partial_t\phi,
}
where $V$ is the voltage in the FQHE and $\epsilon=gV$ for DQM (or
AK). The
effect
of the impurity is then encoded into scattering matrices as
explained in
sections 3 and 4.

\subsec{The Fermi sea}

At $T=0$, in the presence of a voltage,
the ground state is factorized into a Left and a Right Fermi seas.
Moreover, working at zero temperature allows an analytical solution
of the Bethe equations because they become linear.
Restricting to  right movers,  the sea is made of  solitons
filling rapidities
$\theta\in(-\infty,A]$ - this was discussed
in  \flsbig\ to which we refer extensively in the following.
Note that all the results of this section will applly to left movers
as well after substituting antisolitons for solitons (since
the $\partial_t\phi$ term switches sign in \hbulk).

In \flsbig\  the Fermi rapidity, $A$,  is
computed~:
\eqn\fermi{e^A={V\over 2}{G_+(0)\over G_+(i)}.}
In this formula, conventions are such that $\pm V/2$,
is the energy shift of solitons/antisolitons due to the voltage. We
also define  the  ``charge'' of solitons and antisolitons to be $\pm
1$ \foot{It is also convenient to
define their  ``spin'' to be $\pm {1\over g}$. Both
are proportional to the conserved topological  charge of the
hamiltonian. See the
appendix for more details.}. The kernel $G$ is defined by~:
\eqn\ggdef{G_+(\omega)=\sqrt{2\pi \over g}
{\Gamma [-i{\omega\over 2(1-g)}]\over \Gamma
(-i{\omega g\over 2(1-g)})\Gamma({1\over 2}-i{\omega\over 2})}
e^{-i\omega\Delta},}
where~:
$$
\Delta={g\over 2(1-g)}\ln{g}+{1\over 2}\ln(1-g),
$$
and  we define $G_-(\omega)=G_+(-\omega)$.

In the following,  we shall repeatedly encounter the  equation~:
\eqn\geneq{\eqalign{&f(\theta)-\int_{-\infty}^A
\Phi(\theta-\theta')f(\theta')d\theta'
=g(\theta),\cr & \theta\in(-\infty,A[;\quad f=0,\quad \theta>A,
}}
with the (soliton-soliton) phase shift
$\Phi(\theta)={1\over 2i\pi}{d\over d\theta}\ln S(\theta)$~:
\eqn\phideff{\Phi(\theta)=\int_{-\infty}^\infty e^{-i\omega\theta}
{\sinh\pi ({2g-1\over 2(1-g)})\omega
\over 2\cosh{\pi\omega\over 2}\sinh{\pi g\omega\over 2(1-g)}
}{d\omega\over 2\pi}.}
As shown in \flsbig, the function $f$ which solves this  has Fourier
transform~:
\eqn\fourrr{\tilde{f}(\omega)=-{1\over 2i\pi}G_-(\omega)e^{i\omega A}
\int_{-\infty}^\infty {\tilde{g}(\omega')G_+(\omega')e^{-i\omega'
A}\over
\omega'-\omega+i0}d\omega',}
where $\tilde{g}(\omega)=\int_{-\infty}^A
g(\theta)e^{i\omega\theta}d\theta$.
A first  example of equation \geneq\ is provided
by the density of solitons in the Fermi sea which satisfies
(recall
the convention $\hbar=1$)~:
\eqn\denseq{\eqalign{&\rho_+(\theta)-\int_{-\infty}^A
\Phi(\theta-\theta')\rho_+(\theta')d\theta'
={e^\theta\over 2\pi},\cr &
 \theta\in(-\infty,A[;\quad \rho_+=0,\quad \theta>A.
}}
Using \fourrr\ one finds~:
\eqn\denssol{\rho_+(\theta)=
{V\over 2\sqrt{\pi} (1-g)}
\sum_{n=1}^\infty {(-)^n\over n!}{\exp\left[2n
(1-g)(\theta-A-\Delta)\right]
\over \Gamma\left(-ng\right)
\Gamma\left({3\over 2}-n(1-g)\right)}
,}
for $\theta\in (-\infty,A[$,
and $\rho_+=0$ outside the sea. The density $\rho_+$ is discontinuous
at the Fermi rapidity $\theta=A$: one has $\lim_{\theta\to
A^-}\rho_+(\theta)={V\over 4\pi}\sqrt{2g}$.

Similarly, the density of holes of solitons above the Fermi sea
obeys~:
\eqn\holedens{\rho^h_{+}(\theta)={e^\theta\over
2\pi}+\int_{-\infty}^A
\Phi(\theta-\theta')\rho_+(\theta')
d\theta', \ \theta>A.}
This density is the analytic continuation of $\rho_+$ beyond the
Fermi
rapidity. It is sometimes
convenient to introduce a single, analytic quantity $\rho(\theta)$
such that $\rho=\rho_+$
in the Fermi sea, and $\rho=\rho_+^h$ outside the Fermi sea.
Similar equations can be written for the densities of holes of
antisolitons $\rho_-^h$
 and holes of breathers $\rho_n^h$, both quantities which are defined
on the entire
rapidity interval $\theta\in(-\infty,\infty)$.

\subsec{Low energy excitations}

Let us now come to the low energy excitations of the theory
at zero temperature.  As we will show, when there is a voltage,
there is a minimal amount of energy needed to create a
particle or a hole:  The voltage (or bias) introduces a
scale in the theory, and we can look at low (compared with that
scale) energy excitations.

The following processes are the  low energy excitations,
with their associated energies $\epsilon$~:

\item{$\bullet$}  add a soliton: $\epsilon_+(\theta)$,
$\theta\in
]A,\infty)$
\item{$\bullet$}  destroy a soliton (or create a hole):
$\epsilon^h_+(\theta)$,
$\theta\in (-\infty,A[$; $\epsilon_+^h\leq gV$.

The energy of hole and particle excitations can also be addressed
analytically at zero temperature.  For example,
the  excitation energy for  creating a hole in the sea,
$\epsilon^h_+$,obeys the equation~:
\eqn\epseq{\eqalign{&{V\over
2}-e^\theta=\epsilon^h_+(\theta)-\int_{-\infty}^A\Phi(\theta-\theta')
\epsilon^h_+(\theta')
d\theta',\cr & \theta\in(-\infty,A[;\quad
\epsilon_+^h(\theta)=0,\quad,\theta\geq A.
}}
with of course $\epsilon_+^h(\theta)\geq 0$ for $\theta<A$
and $\epsilon^h_+(A)=0$ (this condition being actually
the way of determining the Fermi rapidity \fermi). By differentiating
with respect to $\theta$ one sees that
${d\over d\theta}\epsilon^h_+(\theta)=-2\pi\rho_+(\theta)$. Explicit
solution gives~:
\eqn\epsexp{\epsilon^h_+(\theta)={V\over 2} \sqrt{\pi} g
\sum_{n=1}^\infty {(-1)^{n}\over
n!}{\exp\left[2n(1-g)(\theta-A-\Delta)\right]
\over \Gamma\left(1-ng \right)
\Gamma\left({3\over 2}-n(1-g)\right)}
+Vg.}
the constant term in \epsexp\ is equal to $\epsilon_+^h(-\infty)$,
which can be obtained
straightforwardly from \epseq\ since in that limit this becomes
a simple convolution equation.

For $\theta> A$, $\epsilon_+^h$ as defined above vanishes exactly.
The
excitation energy
to add a soliton above the Fermi sea can be shown to obey~:
\eqn\newexxx{\epsilon_+(\theta)=e^\theta-{V\over 2}-\int_{-\infty}^A
\Phi(\theta-\theta')\epsilon_+^h(\theta')d\theta',}
and thus it is the analytic continuation of $-\epsilon_+^h$ beyond
the
Fermi rapidity.
Using the above determination of $\epsilon_+^h$ one finds
explicitely~:
\eqn\solnewex{\epsilon_+(\theta)=e^\theta -{V\over 2}
+{V\over 4}\sqrt{\pi}\sum_{n=1}^\infty
{(-1)^n\over n!}{1\over\cos
\left(n{\pi(1-g) \over g}\right)}{\exp\left[{2n(1-g) \over
g}(A-\theta-\Delta)\right]
\over\Gamma\left(1-n/g\right)
\Gamma\left({3\over 2}+{n(1-g)\over g}\right)}
.}
As before, we define an analytic quantity $\epsilon$
which is equal to $-\epsilon_+^h$ in the sea and $\epsilon_+$ outside
the sea.

As an example, let us prove \newexxx.
Call $\rho$ the density of solitons
 in the sea and $\rho_p$ the density of solitons
above the sea (often referred to in what follows as particles).
 If we add a few solitons at rapidities $\theta_1,\ldots,
\theta_N$, we induce a change of density above the sea,
$\delta\rho_p(\theta_p)=
{1\over L}\sum_i \delta(\theta_p-\theta_i)$, as well as a change of
densities
in the sea through the Bethe equations. The quantity we are looking
for obeys, by definition~:
\eqn\junki{\delta E=L\int_A^\infty
\epsilon_+(\theta_p)\delta\rho_p(\theta_p)
d\theta_p,}
while on the other hand~:
\eqn\junkii{\delta E= L\int_{-\infty}^A \left(e^\theta -{V\over
2}\right)
\delta\rho(\theta)d\theta +L\int_A^\infty \left(e^{\theta_p}-{V\over
2}\right)\delta
\rho_p(\theta_p)d\theta_p.}
{}From the Bethe equations one has~:
\eqn\bethshift{\delta\rho(\theta)=\int_{-\infty}^A
\Phi(\theta-\theta')\delta\rho(\theta')
d\theta'+\int_A^\infty \Phi(\theta-\theta_p)\delta
\rho_p(\theta_p)d\theta_p.}
Let us write the solution of the general equation \geneq\ as~:
\eqn\kerker{
f(\theta)=\int_{-\infty}^A [\delta(\theta-\theta')+
L(\theta,\theta')]g(\theta')d\theta',}
where the kernel $L$ is a symmetric function
of its two arguments, and its exact expression  is not needed in what
follows. Solving \bethshift\
for $\delta\rho$ and replacing in \junkii\ leads to~:
\eqn\morjunk{\epsilon_+(\theta)=e^\theta-{V\over 2}+
\int_{-\infty}^A \Phi(\theta-\theta')\int_{-\infty}^A
[\delta(\theta'-\theta'')+L(\theta',\theta'')]
\left(e^{\theta''}-{V\over 2}\right)d\theta''.}
Equation \newexxx\ follows then from \epseq.

The corresponding analysis for the momentum presents a subtlety. The
reason is,
that the momentum of excitations does not
vanish at the Fermi velocity, as can easily be seen with the example
of free fermions ($g=1/2$). Therefore, one has to be very careful
with what
happens
right at the Fermi surface - this difficulty did not appear for
the energy because $\epsilon$ vanishes at the Fermi surface anyway.
We will discuss the
dressed momentum in more details in the next section. Of course,
the final result can be predicted on physical grounds:
excitations have to be  {\bf relativistic}, as they obviously are
in the $g=1/2$ case.
Indeed, adding a potential $V$
amounts (in the bulk)
to shifting $\partial_t\phi$ by a constant, and this does not change
the
fact that
excitations have a  relativistic spectrum - more precisely, for
excitations that
have as many particles as holes, momentum $=$ energy, while
for non-neutral ones, momentum $=$ energy  $+$ constant.

In the following we define
$p_+=\epsilon_+$ and $p_+^h=\epsilon_+^h$

While at  zero temperature,
there are only solitons, antisolitons
as well as breathers do contribute to the excitation
spectrum, as we now discuss.

\subsec{Other excitations}

Other excitations occur at a finite energy above the
ground state (for $g<1$) and are obtained by the following
processes~:

\item{$\bullet$}  add a antisoliton: $\epsilon_-(\theta)$,
$\theta\in (-\infty,\infty)$; $\epsilon_-\geq (1-g)V$.
\item{$\bullet$}  add an n-breather: $\epsilon_n(\theta)$, $\theta\in
(-\infty,\infty)$; $\epsilon_n\geq ngV$

For
adding a antisoliton one finds that the excitation energy is
$\epsilon_-(\theta)=V+\epsilon_+(\theta)$ if $\theta>A$ and
$\epsilon_-(\theta)=V-\epsilon_+^h(\theta)$
if $\theta<A$. In particular, since the maximum energy for a hole is
$\epsilon_+^h(-\infty)=gV$,
we see that the threshold to add a soliton is $(1-g)V$. This implies
in particular
that the low-energy processes studied above must have an energy
$\omega<<(1-g)V$,
and that the limit $g=1$ is highly singular in this approach.
Similarly,
these low energy processes must have an energy $\omega<<gV$, and the
limit $g=0$ is also singular.

Different physical processes can therefore occur
in the excitations, which have different thresholds. The
 structure of these thresholds is quite intricate for $g$
arbitrary. In what follows,
we shall restrict to the usual case $g={1\over t}$, $t$ an integer.
Then the thresholds
occur at energy values ${p\over t}V$, ie all multiples of $gV$.

\subsec{Charge dressing}

It is also interesting to consider the charge of excitations. Suppose
we
destroy a soliton at rapidity $\theta$ in the sea. We define the
dressed
charge  $q_+^h(\theta)=2{d\over dV}\epsilon_+^h$ \foot{This
definition
is adequate only near the Fermi rapidity.}.
It
is
therefore the solution of the  equation
similar to $\epsilon_+^h(\theta)$ \epseq, but with the left hand side
${V\over 2}-e^\theta$
replaced by the opposite of the charge of the bare solitons~:
\eqn\drch{-1=q_+^h(\theta)-\int_{-\infty}^A\Phi(\theta-\theta')
q_+^h(\theta')d\theta'.}
This is actually related to other quantities we already computed in
the sea:
${V\over 2}q_+^h(\theta)=-2\pi\rho_+(\theta)+\epsilon_+^h(\theta)$,
from
which
follows the renormalized charge of the excitations at the Fermi
surface~:
\eqn\rench{q_+^h(A)=-q_+(A)-=-\sqrt{2g},}
where we used above values of $\rho_-(A)$. In \rench, $q_-$ is
similarly
the value of the dressed charge for an antisoliton added above the
Fermi surface. Of course, breathers have a vanishing dressed charge.

\newsec{Physics at low energy.}

\subsec{Dressed scattering}

While excitations have a  relativistic
dispersion relation, relativistic invariance
is broken by the existence of the energy scale $e^A$.
The next question is then, what sort of relativistic
theory is that? To find out, we must determine the corresponding $S$
matrix. First, some technicalities.  Introduce the functional
$\hat{K}$ such that~:
$$
\hat{K}\bullet f(\theta)=\int_{-\infty}^A
\Phi(\theta-\theta')f(\theta')d\theta'
$$
and similarly~:
$$
\hat{I}\bullet f(\theta)=f(\theta)
$$
Then \geneq\ reads $(\hat{I}-\hat{K})\bullet f=g$, and the general
solution \kerker\ reads in those terms~:
$$
(\hat{I}+\hat{L})(\hat{I}-\hat{K})=\hat{I}
$$
or equivalently $(\hat{I}+\hat{L})\hat{K}=\hat{L}$. This means in
turn that~:
$$
\eqalign{\int_{-\infty}^A
[\delta(\theta-\theta')+L(\theta,\theta')]d\theta'
\int_{-\infty}^A \Phi(\theta'-\theta'')f(\theta'')d\theta''\cr
=\int_{-\infty}^A L(\theta,\theta')f(\theta')d\theta'.\cr}
$$
Since this is true for any function $f$ we deduce the identity~:
\eqn\niceident{
\int_{-\infty}^A
[\delta(\theta-\theta')+L(\theta,\theta')]
\Phi(\theta'-\theta'')d\theta'=
L(\theta,\theta'').}

Let us now consider the quantization equation allowing a few
particles and holes. One has
\eqn\newversi{2\pi(\rho+\rho^h)(\theta)=e^\theta+2\pi\int_{-\infty}^A
\Phi(\theta-\theta')\rho(\theta')d\theta'+\int_A^\infty
\Phi(\theta-\theta')\rho_p(\theta')d\theta'.}
Let us now transform this equation to make sense in the dressed
theory
\ref\KR{A. N. Kirillov, N. Yu
Reshetikhin, J. Phys. A20 (1987) 1587.}.
 In this theory, the excitations are still particles above the sea,
but they
 are holes
in the sea. Therefore, the dressed equations must have on the right
hand side
 not $\rho$, but rather $\rho_h$. After some manipulations
using the foregoing technical indentities, we obtain
\eqn\newversii{2\pi(\rho+\rho^h)(\theta)=-
2\pi\rho_+-2\pi\int_{-\infty}^A
L(\theta,\theta')\rho^h(\theta')d\theta'+2\pi\int_A^\infty
L(\theta,\theta')\rho_p(\theta')d\theta'.}

The first term in this equation must be identified with the
derivative of the dressed momentum of holes, ie one has ${d\over
d\theta}p_+^h(\theta)=-2\pi\rho_+(\theta)$. But it is easy to see
that this equals in turn ${d\epsilon_+^h\over d\theta}$, proving as
claimed in the
previous section that hole excitations have a dispersion relation of
the form
momentum $=$ energy + constant. The two other terms must be
identified
with the dressed scattering, and therefore we have \foot{In fact, the
following relations hold only for $\theta<\theta'$, since the
relation between the S matrix and the phase shift depends
on which particle has the largest rapidity}.
\eqn\newversiii{\eqalign{{1\over 2i\pi}{d\over d\theta'}\ln
S_{hh}(\theta,\theta')=L(\theta,\theta')\cr
{1\over 2i\pi}{d\over d\theta'}\ln
S_{hp}(\theta,\theta')=-L(\theta,\theta').\cr}}
Similarly we have
\eqn\newversiv{\eqalign{{1\over 2i\pi}{d\over d\theta'}\ln
S_{pp}(\theta,\theta')=L(\theta,\theta')\cr
{1\over 2i\pi}{d\over d\theta'}\ln
S_{ph}(\theta,\theta')=-L(\theta,\theta'),\cr}}
together with ${d\over
d\theta}p_+(\theta)=2\pi\rho_+^h(\theta)={d\over
d\theta}\epsilon_+(\theta)$, proving that particle excitations
are also relativistic.

To determine completely the S-matrix, one has to
find out the constants of integration. This is a question
that is not directly answered by the Bethe ansatz. Usually,
these constants would be determined by using relativistic invariance,
ie requiring
that the full S matrix depends only on the difference
of rapidities \korbook. Here however, relativistic invariance
is broken, and the S matrix has a more complicated dependence on
these
rapidities. To compute the constants,
one needs to reformulate the equivalent of crossing and unitarity
in the dressed theory with broken relativistic invariance. Here
we shall contend ourselves with an analysis
of the low energy excitations, near the Fermi rapidity. As $\theta\to
A$, the
energy scale $V$ becomes unimportant because $\epsilon<<V$,
$\epsilon/V\to 0$,
and relativistic invariance is restored.
Then, $L(\theta,\theta')$  can be approximated by  $L(A,A)$, and
integration of \newversiii\ and \newversiv\ combined with unitarity
leads to the simplest solution:
\eqn\renslim{\eqalign{{\cal
S}_{ph}(\theta_p,\theta_h)&=-\exp\left[2i\pi
L(A,A)(\theta_p-\theta_h)
\right]\cr
{\cal S}_{hp}(\theta_h,\theta_p)&=-\exp\left[2i\pi
L(A,A)(\theta_h-\theta_p)
\right]\cr
{\cal S}_{pp}(\theta_p,\theta'_p)&=-\exp\left[-2i\pi
L(A,A)(\theta_p-\theta'_p)
\right]\cr
{\cal S}_{hh}(\theta_h,\theta'_h)&=-\exp\left[-2i\pi
L(A,A)(\theta_h-\theta'_h)
\right]\cr.}}
The overall minus sign here is necessary to ensure that no two
particles
or holes can coincide \foot{The sign of S at identical rapidities
is a subject of some discussion, and largely dependent on the
way the S matrix is defined. For us, S is the object appearing in the
Fateev Zamolodchikov algebra.}. Let us stress that \renslim\ is
, due to the problem of integration constants, partly  a conjecture,
which we will check carefully
in particular in the next subsection. For a more complete discussion
of \renslim,
see the next paper \ref\LSnext{F. Lesage, H. Saleur, in preparation.}

For low energy excitations above the sea, this effective theory
represented by \renslim\ is
simply a free fermion theory: the additional phase shifts being of
the
form $\exp[i\ cst (\theta- \theta')]$
amount to a simple gauge transformation and can be gauged away
(alternatively,
form-factors for \renslim\ are free fermion form-factors, up to
phases that cancel out in the correlators of interest). Without the
sea, if one looks
at say the bare $S$ matrix of solitons, it  does not have a well
defined limit at very low energy, ie when
both rapidities approach minus infinity, because by relativistic
invariance it depends on  the ratio of these energies
$e^{\theta_1}/e^{\theta_2}$. The
role of the  sea is to give  the dressed S matrix a well defined
limit at very low energy, ie
when both rapidites approach $A$. It is then
very natural that the limiting theory should be a theory of free
fermions.

In subsequent computations, it will be useful to introduce the shift
function.
For example,  suppose we add a soliton above the sea
at rapidity $\theta_p$ and create a hole in the sea at rapidity
$\theta_h$. This induces a shift of the rapidities
in the sea: a rapidity equal to $\theta_\alpha$ initially becomes
$\theta_\alpha
+\delta^{(2)}\theta_\alpha$ with the conditions~:
\eqn\conds{\eqalign{I_\alpha=&{1\over 2\pi}Le^{\theta_\alpha}+{1\over
2i\pi}
\sum_\beta \ln S(\theta_\alpha-\theta_\beta)\cr
I_\alpha=&{L\over
2\pi}e^{\theta_\alpha+\delta^{(2)}\theta_\alpha}+{1\over 2i\pi}
\sum_\beta \ln
S(\theta_\alpha-\theta_\beta+\delta^{(2)}\theta_\alpha-\delta^{(2)}
\theta_\beta) \cr
&+{1\over 2i\pi}\ln S(\theta_\alpha-\theta_p)-{1\over 2i\pi}\ln
S(\theta_\alpha-\theta_h).\cr}}
As usual
we define the shift function by
$L\rho_+(\theta)\delta^{(2)}\theta\equiv F(\theta|\theta_p,
\theta_h)$.
By standard manipulations, one finds
the equation obeyed by the shift:
\eqn\shifti{{1\over 2i\pi}[\ln S(\theta-\theta_h)-\ln
S(\theta-\theta_p)]=F(\theta|\theta_p,\theta_h)-
\int_{-\infty}^A
\Phi(\theta-\theta')F(\theta'|\theta_p,\theta_h)d\theta'.}
A formal solution of this equation
follows as~:
\eqn\form{F(\theta|\theta_p,\theta_h)=
\int_{\theta_h}^{\theta_p}L(\theta,\theta')d\theta.}

\subsec{The noise at low energy in a  pure Luttinger liquid}

To verify the consistency of our low-energy approach, let
us consider the current noise in a Luttinger liquid with a voltage
and
in the absence of impurity. Let us work within
the effective theory of low energy excitations. In doing so,
we forget completely about the Fermi sea and concentrate on the
excitations. Let
us write their energies and momentum $e=p<<V$.  It is convenient here
to
introduce new rapidities such that for particles
$\epsilon_+(\theta_p)\equiv e^{\beta_p}$ and
for holes $\epsilon_+^h(\theta_h)\equiv e^{\beta_h}$. Consider then
the
renormalized current operator where the vacuum expectation value has
been subtracted.
For a free Fermion theory one has~:
\eqn\ff{\eqalign{<0|:\partial_z\phi:(0,0)|\beta_1,\beta_2>_{-+}
=c e^{\beta_1/2}e^{\beta_2/2}\cr
<0|:\partial_z\phi:(0,0)|\beta_1,\beta_2>_{+-} =-c
e^{\beta_1/2}e^{\beta_2/2},\cr}}
where all the particles are assumed to be right moving, we changed
notation to call a particle $+$
and a hole $-$, and $c$ is a constant to be determined.
 To fix the normalization $c$
 of this form-factor, let us consider the charge of the
one particle state
\eqn\charge{\eqalign{{}_+<\beta_1 &|\int_{-\infty}^\infty dx
:\partial_x\phi:(x,t)|\beta_2>_+\cr &=\int_{-\infty}^\infty dx
<0|:\partial_z\phi:(0,0)|\beta_1-i\pi,\beta_2>_{-+} \exp[
i(x-t)(e_2-e_1)]\cr
&=-2\pi ic e^\beta{\delta(\beta_1-\beta_2)\over de/d\beta}\cr
&=2\pi \sqrt{2g}\delta(\beta_1-\beta_2),\cr}}
where the last equality is imposed using the foregoing dressed charge
computation.
It follows that $c=i\sqrt{2g}$. Hence at coupling $g$, all what
differs
from the $g=1/2$ case is a renormalization
of the two-particle  form factor by $\sqrt{2g}$. The computation of
the correlator
is then the same as for the free fermions, up to a renormalization by
$(\sqrt{2g})^2$
since the two point function of the current involves the squares of
form-factors.
We find then the  noise at low energy~:
\eqn\noisepure{S(\omega)\approx {g\over 2\pi}|\omega|,}
in agreement with the well known exact result.

Let us stress that this little computation is actually a non trivial
check. In general, and for vanishing voltage, the normalization of
the
two particle form-factor can be determined by imposing the value of
the
charge. When one computes the contribution of this two particle
form-factor to
the two point function of the current, it appears that some
contributions
are missing, due to higher  form-factors: in other words higher
form-factors
are needed without voltage because the charge of excitations is $1$
while the two point function has amplitude $2g$. With a voltage,
things are completely different, because the dressed charge is
$\sqrt{2g}$, so the (free)
two-particle form-factor is sufficient to reproduce the two-point
function
amplitude. Let us stress also that there is no way
one can reproduce the $V=0$ case by taking a ``limit'' $V\to 0$. In a
massless
theory, there is no such thing as a low energy scale.

\newsec{Thresholds and potential singularities}

As explained earlier, the interacting theory
presents a series of thresholds : physical processes
become allowed or forbidden when $\omega$ crosses one of the values
$\omega=ngV$. Without an impurity, none of these values does actually
lead to a singularity in the noise, since the formula $S={g\over
\pi}|\omega|$
holds. What takes place are  very special cancellations, which
generally will be spoilt by the impurity.   A simple example to see
the phenomenon
is the free fermions $g=1/2$. Consider the noise close to the
threshold $\omega=gV={V\over 2}$. For $\omega< {V\over 2}$, the only
allowed physical process is the creation of a particle hole pair.
 Using that the form factor is
$e^{\beta_1/2}e^{\beta_2/2}$ one can write~:
$$
\eqalign{\int_0^{V/2}\int_{0}^\infty de_1de_2 \
\delta(\omega-e_1-e_2)&=
\int_0^\omega de_1=\omega,\ \omega<V/2\cr
&=\int_0^{V/2} de_1=V/2,\ \omega>V/2,\cr}
$$
where we used that for any $e_1$ in the interval $[0,V/2]$,
$\omega-e_1>0$ in the second  case.
Of course when $\omega>V/2$, another process is possible beside
exciting
a particle from the sea, it is the creation of a pair
soliton-antisoliton. The threshold
for creating an antisoliton is at $V/2$, and the form factor is the
same so one has then~:
$$
\int_{V/2}^\infty \int_0^\infty de_1de_2 \ \delta(\omega-e_1-e_2)=
\int_{V/2}^\omega de_1={V\over 2}-\omega,
$$
and this second process adds up to the previous one to reproduce the
$\omega$ dependence
at all values of $\omega$. Hence there is no singularity because the
term
involving solitons produces
an analytic continuation  of the term involving a hole.

Since we know there is no singularity in the noise, in the absence of
impurity, for any
frequency, this means that
similar cancellations must take place between various contributions
around a given threshold.
For instance, $(1-g)V={t-1\over t}V$ is the threshold at which
soliton
creation is possible, while it is the maximum energy one can reach by
creating $(t-1)$ holes in the sea, etc.

Let us now investigate in more details the potential singularity at
$\omega_c=gV$, assuming $g<{1\over 2}$. At this
frequency, the process involving a pair particle hole saturates
while the process involving creation of a (first) breather above the
Fermi
sea kicks in.
Let us consider in more details the process involving a breather
above the Fermi sea.
Once again, we will restrict to low energies that is to frequencies
such that $|\omega-
\omega_c|<<\omega_c$. In that limit, the breathers are created at
rapidities close to
$-\infty$, and in all the computations  determining  the phase shifts
and the dressed S-matrix,  one can replace
the integrals on $[-\infty,A]$ by integrals over the whole real axis.
The renormalized
breather-breather S-matrix is then found to be~:
\eqn\rensbbbb{{1\over 2i\pi}{d\over d\theta}\ln {\cal S}_{bb}=
\Phi_{11}+
 {\Phi_1^2\over 1-\Phi},}
(where the right hand side is understood in terms of Fourier
transforms,
$\Phi_{11}$ and $\Phi_1$ are defined like $\Phi$ as the logarithmic
derivatives of the breather-breather and breather-soliton
S-matrices), and this
simply reproduces $\Phi_{11}$ up to a rapidity renormalization.
Hence, the renormalized
scattering theory
for breathers close to $-\infty$ is insensitive to the voltage, and
behaves like an ordinary sine-Gordon model.  Setting~:
\eqn\enparam{\epsilon_b(\theta)=gV+e^{\beta},}
 the form factor of the current can depend only on $e^\beta$, and
by dimensional analysis, must  simply be proportional to
$e^{\beta}$.
Hence
 the leading  contribution
to the noise without impurity of the breather term is  ~:
\eqn\brthcontr{{\cal A}_b\int_{-\infty}^\infty e^{2\beta}
\delta\left(\omega-\omega_c-e^{\beta}\right)d\beta
= {\cal A}_b\ (\omega-\omega_c), \omega>\omega_c; \ \
0, \omega<\omega_c}
where ${\cal A}_b$ is an  amplitude we did not determine here.
Similarly let us consider the particle hole term. Calling
$f(\theta_1,\theta_2)$
the form factor of the current between the Fermi sea and a state with
a hole at $\theta_1<A$
and an added particle at $\theta_2>A$, the contribution to the noise
is proportional to~:
\eqn\uglyyy{\int_{-\infty}^A d\theta_1\int_A^\infty d\theta_2
\vert f(\theta_1,\theta_2)\vert^2 \
\delta\left[\omega-\epsilon_+(\theta_2)
-\epsilon_+^h(\theta_1)\right].
}
Performing the $\theta_2$ integral gives, for $\omega<\omega_c$~:
\eqn\uglyyyy{\int_{(\epsilon_+^h)^{-1}(\omega)}^A
\vert f(\theta_1,\theta_2)\vert^2
{1\over\dot{\epsilon}_+(\theta_2)}
d\theta_1,\quad
\theta_2=(\epsilon_+)^{-1}(\omega-\epsilon_+^h(\theta_1)),}
while for $\omega>\omega_c$, the integral runs from $-\infty$ to
$A$ instead,
with  $\epsilon_+^h(-\infty)=\omega_c$. Hence the
contributions of the particle hole term  below and above $\omega_c$
differ by the integral from
$-\infty$ to
$(\epsilon_+^h)^{-1}(\omega)$ of the same integrand
as in \uglyyyy.

Now, for $\omega\approx\omega_c$,
the values of $\theta_1$ and $\theta_2$ are very far apart:
 $\theta_2\approx A$, a
finite rapidity, while $\theta_1$ approaches the
value $-\infty$.  At leading order,
$\vert f(\theta_1,\theta_2)\vert^2/\dot{\epsilon}_+(\theta_2)$
must then factorize into a function of $\theta_1$:  $f(\theta_1)$,
and a function of $A$.
Setting~:
\eqn\uglyiyiy{\epsilon_+^h(\theta_1)=\omega_c-e^{\beta_1},}
we trade $\theta_1$ for $\beta_1$. By dimensional analysis,
and one has $f(\theta_1)\propto e^{\beta_1/2}$,
so introducing
 the variable $x=e^{\beta_1}$, the contribution to the noise without
impurity of the particle hole term
part can be rewritten, in addition to a regular term, as~:
\eqn\notsougly{{\cal A}_{ph}\int_{\omega}^{\omega_c} dx={\cal
A}_{ph}\ (\omega_c-\omega), \omega<\omega_c}
${\cal A}_{ph}$ being another amplitude. The  two processes
which ``cross'' at $\omega_c$ behave both linearly in
$\omega-\omega_c$ at leading order (there are no
other processes near  $\omega_c$, and we discuss only the leading
singularities). Since the
noise has no singularity in the absence of impurity, this means
that \brthcontr\ must be the analytic continuation of \notsougly,
that
is ${\cal A}_b={\cal A}_{ph}$.

For $\omega_c=ngV, n\geq 2$, there are in addition background
processes,
that is processes present on both sides of $\omega_c$. We shall
assume
that such processes do not have singular behaviours - equivalently
that the
form factors of the renormalized theory are regular around
$\omega_c$.

Then, for
 $\omega_c=2gV$, we have the  one second-breather process
for $\omega>\omega_c$,
 the two particles - two holes process for $\omega<\omega_c$:
the previous analysis thus generalizes
 to this case.

 For $\omega_c=ngV, n>2$, more than two processes usually cross,
making the
analysis
more difficult: for instance for $\omega=3gV$, the one third-breather
process
together with the three one-breather process are possible for
$\omega>\omega_c$.
The absence of singularity for the noise of the pure system therefore
does not
completely constrain the amplitudes as was the case for
$\omega_c=gV$.

 We now apply the above considerations to the two problems of
physical
interest.

\newsec{Properties  of the response function in the
two-state problem.}

\subsec{Low frequency behaviour.}

As a first application, we consider the low energy behaviour
of the response function in the two-state problem with a bias $V$.

In the presence of a boundary interaction we have several energy
scales: $T_B,\omega,V$.
The function ${\chi''(\omega)\over\omega}$ can be written as ${1\over
T_B^2}$ times
a function $F\left({\omega\over V},{ T_B\over V}\right)$. We will be
able to
use the foregoing low-energy  scattering theory of excitations above
the Fermi
sea provided we restrict to $\omega<<V$: that is only the dependence
of $F$ on the first variable will be accessible. Observe this is
enough
to obtain the value $\lim_{\omega\to 0} F$.

At low energy, the only physical processes
involve R-solitons and L-antisolitons. The boundary interaction is
then
fully characterized by the reflection
matrix $R^+_-$ - in this paragraph we shall omit the isotopic indices
from now on. In the presence of the boundary, the ground state
becomes~:
\eqn\gstvolt{
||\theta_1,\cdots , \theta_n>.
}
It is  a mixture of left and right particles
\LSS(eg $||\theta>=|\theta>^R+
R|\theta>^L$). The  rapidities are the same
as was discussed in section 2 because there is no LR scattering.

The low energy results are obtained by making a particle-hole
pair
 $\theta_p, \theta_h$,
with the  remaining rapidities being shifted.  For the
current-current correlator, this gives ~:
\eqn\ins{\eqalign{
&<\theta_n,\cdots,
\theta_1||J_R||\tilde{\theta_1},...,\hat{\theta}_h,...
\tilde{\theta_n},\theta_p>
<\theta_p,\tilde{\theta_n},...,\hat{\theta}_h,...,\tilde{\theta_1}||
J_L||\theta_1,...,\theta_n>= \cr &
{}^{R...R}<\theta_n,\cdots,
\theta_1|J_R|\tilde{\theta_1},...,\hat{\theta_h},...
\tilde{\theta_n},\theta_p>^{R...R} \times \cr &
{}^{L...L}<\theta_p,\tilde{\theta_n},...,
\hat{\theta_h},...,\tilde{\theta_1}|
J_L|\theta_1,...,\theta_n>^{L...L}
R^*(\theta_p) \prod_{i\neq h} R^*(\tilde{\theta_i}) \prod_{i}
R(\theta_i)
}}
where the hat denotes the omitted (hole) rapidity.
Observe that \ins\ involves the same matrix elements
as for the excitations in the bulk, to which the analysis
of section 3 applies.  As for the $J_LJ_L$ and $J_RJ_R$
correlators, they are not affected by the boundary interaction.
Using the fact that
$R^*=R^{-1}$ and inserting $R^*(\theta_h)R(\theta_h)=1$, we
find that we can express these reflection matrices in terms
of a renormalised reflection matrix for the particles and holes~:
\eqn\whouah{R^*(\theta_p)R(\theta_h)\to \exp\left\{
\int _{-\infty}^A [F(\theta|\theta_h)-F(\theta|\theta_p)]{d\over
d\theta}\ln R(\theta)
\right\}R^*(\theta_p)R(\theta_h).}
It is then convenient to introduce a renormalized reflection matrix~:
\eqn\renrdef{{\cal R}(\theta)=R(\theta)\exp\left[\int_{-\infty}^A
 F(\theta'|\theta){d\over d\theta'}
\ln R(\theta')\right].}
Using the same line of arguments as in \LSS , one can write rerwrite
$\chi''$
in terms of the correlator of the current operator. The latter can
then
be expressed through form-factors. One finds:
\eqn\gri{\chi''(\omega)=-{1\over 2g\pi^2\omega^2}\hbox{Re }
\int_{-\infty}^{\ln\omega}d\beta_2 d\beta'_2\left[ {\cal
R}^*(\theta_2){\cal
R}(\theta'_2)-1\right]e^{\beta_2}e^{\beta'_2}
\delta(\omega-e^{\beta_2}-e^{\beta'_2}),}
where we have defined the $\theta\to \phi$ correspondence through
$\epsilon_+^h(\theta_2)=e^{\beta_2}$ and
$\epsilon_+(\theta'_2)=e^{\beta'_2}$.
To clarify this, let us write the $R$ part explicitely~:
$$
R(\beta'_2)={e^{\theta'_2}+iT_B\over e^{\theta'_2}-iT_B}={e^A+\alpha
e^{\beta'_2}+iT_B\over
e^A+\alpha e^{\beta'_2}-iT_B},
$$
where~:
\eqn\alphdef{
\alpha= { e^A\over 2\pi\rho(A)}={G_+(0)\over \sqrt{2g}G_+(i)},}
(here we used $2\pi\rho(A)={d\epsilon\over d\theta}\vert_A$) and
similarly~:
$$
R^*(\beta_2)={e^A-\alpha e^{\beta_2}-iT_B\over
e^A-\alpha e^{\beta_2}+iT_B},
$$
where the additional minus sign arises from the different
hole/particle parametrization. In the foregoing equations, $T_B$
is a renormalized coupling, related with the bare coupling $\lambda$
by
 \flsbig\
\eqn\rencoupliii{\lambda\kappa^{-g}= 2\sin (\pi g){2^g\over 4\pi}
\Gamma(g)
\left(2ge^{-\Delta}{G_+(i)\over G_+(0)}T_B\right)^{1-g},}
where the expressions for $G_+$ and $\Delta$ are given in section
2.1, $\kappa$
is a cut-off. For simplicity, we usually use the variable $T_B$
in the sequel.

The integral in \gri\ can  then  rewritten, setting
$e^{\beta'_2}\equiv x$~:
\eqn\forsecd{
\int_0^\omega [{\cal R}(x){\cal R}^*(x-\omega)-1] dx,}
(note the dependence on $x-\omega$ instead of $\omega-x$ for the same
reason as above). Obviously this vanishes when $\omega=0$. Because
$|R|^2=1$, the
term linear in $\omega$ vanishes too. At second order
one would have~:
$$
\omega\int_0^\omega {d\over dx}\ln{\cal R} dx
$$
and this will not contribute when we take the real part because $R$
is
a pure phase.   The imaginary part of this expression, though,
contributes
to the real part of the static spin-spin succeptibility
$\chi=\chi'+i\chi''$
\LSS .  Its expression is given by~:
\eqn\kipp{
\lim_{\omega\rightarrow 0}\chi'(\omega)=\chi_0={i\alpha\over 2\pi^2
g}
\left(e^{-\theta}\left.{d\over d\theta}\ln {\cal
R}(\theta)\right|_{\theta=A}\right).
}

For $\chi''$,
we find that the first non trivial term goes like $\omega^3$
as expected. Collecting all terms one finds~:
\eqn\mainmain{\lim_{\omega\to 0}
{\chi''(\omega)\over\omega}=-{\alpha^2\over 4 g\pi^2}
 \left(e^{-\theta}\left.{d\over d\theta}\ln {\cal
R}(\theta)\right|_{\theta=A}\right)^2.}
 From these two previous expressions we can prove
Shiba's relation in the presence of a bias~:
\eqn\shib{
\lim_{\omega\rightarrow 0} {\chi''(\omega)\over \omega}
=\pi^2 g \chi_0^2,
}
which is exactly the same as the one without bias\foot{
This relation  differs from the usual
$\lim_{\omega\rightarrow 0} \chi''(\omega)/\omega=2\pi g \chi_0^2$
because of a different convention for Fourier transforms
and spin normalization.}.

In particular for $g=1/2$ there is no renormalization due to the sea,
${\cal R}(\theta)
=R(\theta)={e^\theta+iT_B\over e^\theta-iT_B}$ and thus, using
${V\over 2}=e^A$ in that case~:
\eqn\mainapp {\lim_{\omega\to 0}
{\chi''(\omega)\over\omega}={2\over\pi^2} {T_B^2\over\left(
{V^2\over 4}+T_B^2\right)^2},}
in agreement with the exact result \ref\sassweissstoul{M. Sassetti,
U. Weiss,
Phys. Rev. A41 (1990) 5383.}.

Let us rewrite \mainmain\ in a more explicit form~:
\eqn\mainmainnew{\lim_{\omega\to 0}
{\chi''(\omega)\over\omega}={\alpha^2\over 4g\pi^2}
e^{-2A} \left({1\over \cosh(A-\theta_B)}+
\int_{-\infty}^A \left.{d\over d\theta}
F(\theta'|\theta)\right|_{\theta=A}
{1\over\cosh(\theta'-\theta _B)}d\theta'\right)^2.}
Let us first investigate the behaviour as $V/T_B\to 0$. At leading
order the term in the bracket can be rewritten as~:
$$
\eqalign{
&{2\over T_B}\left(e^A+\int_{-\infty}^A e^{\theta'} {d\over
d\theta}\left. F(\theta'|\theta)
\right|_{\theta=A}\right)\cr
&={2\over T_B}\left(e^A+\int_{-\infty}^A e^\theta
L(\theta,A)d\theta\right)\cr
&={4\pi\over  T_B}\rho(A).\cr}
$$
where we used \denseq, \kerker, \form. Hence, at very small voltage
one
has~:
\eqn\smallv{\lim_{\omega\to 0}
{\chi''(\omega)\over\omega}={\alpha^2\over 4g\pi^2} {1\over T_B^2}
\left({4\pi\rho(A)\over  e^A}\right)^2={1\over g(\pi T_B)^2}.}
This result is identical with the value of $\lim_{\omega\to 0}
{\chi''(\omega)\over\omega}$
at {\bf vanishing voltage} \LSS. This proves that the limit
$\omega\to
0$ and $V\to 0$
commute, as seems clear from the NRG method \costi. Notice
that a priori this
result is non trivial, the structure of excitations being
very different at vanishing and non-vanishing voltage in a Luttinger
liquid.

To obtain more insight in the behaviour of  $\lim_{\omega\to 0}
{\chi''(\omega)\over\omega}$, some transformations are useful,
which are related with the standard Bethe ansatz computation
of the susceptibility.

\subsec{Static succeptibility by the Bethe ansatz.}

Since we find a relation between
$\lim_{\omega\to 0}{\chi''(\omega)\over\omega}$ and
the static susceptibility $\chi_0$,
we can perform a crucial check of our approach
since $\chi_0$ can be computed by other means.
Indeed, the total spin succeptibility
is simply a second derivative of the
free energy with respect to $V$, and the free
energy can be computed directly by the Bethe-ansatz.

By using standard manipulations for boundary theories
(see eg \ref\FWS{P. Fendley, H. Saleur, N.P. Warner, Nucl. Phys. B340
(1994) 577.}), we find
for the impurity part of the free energy at zero temperature~:
\eqn\impfree{
{\cal F}_{imp}=-{1\over 2\pi} \int_{-\infty}^A d\theta\
{d\over d\theta}\ln R(\theta-\theta_B) \epsilon_+^h(\theta).
}
 The
impurity susceptibility  is then
given by  $\chi_0\equiv -{1\over 2\pi}{d^2{\cal F}_{imp}\over
d(\epsilon/2)^2}$. It reads generally~:
\eqn\geng{
\chi_0=
{1\over g^2\pi^2}\left(
{d\over d\theta}\ln R(A-\theta_B) {1\over V}
\left.{\partial\epsilon_+^h(\theta)\over \partial V}
\right|_{\theta=A}+\int_{-\infty}^A d\theta
{d\over d\theta}\ln R(\theta-\theta_B)
{\partial^2\epsilon_+^h(\theta)\over \partial
V^2}
\right) ,
}
where we used the correspondence $\epsilon=gV$. Since
${d\over d\theta}\ln R(\theta)={1\over\cosh\theta}$, we see that this
is
almost in the same form as the previous expression \mainmainnew\ for
$\chi_0$.
The exact correspondence
can be established using~:
$$
\eqalign{
{\partial\epsilon_+^h(\theta)\over\partial V}&=g+
{(\epsilon_+^h(\theta)-Vg)\over V}
-{\partial\epsilon_+^h(\theta)\over \partial \theta}
{\partial A\over \partial V} \cr &=
{\epsilon_+^h(\theta)\over V}+{2\pi\over V}
\rho_+(\theta),
}
$$
from which it follows that~:
\eqn\frediii{\left.{\partial\epsilon_+^h(\theta)
\over\partial V}\right|_{\theta=A}={\sqrt{2g}\over 2}.
}
Similarly, simple manipulations using the $L$ operator of sections 2
and 3 lead to the identity~:
\eqn\frediv{
{\partial^2\epsilon_+^h(\theta')
\over\partial V^2}={\sqrt{2g}\over 2}{1\over
V}L(\theta',A)=
{\sqrt{2g}\over 2}{1\over V}\left.{\partial
F(\theta'|\theta)\over\partial\theta}\right|
_{\theta=A},}
completing the proof of the identity.

{}From this discussion the static
susceptibility, after a few manipulations, reads ~:
\eqn\gengsimp{
\chi_0=
{2 T_B\over \pi g^2 V^2}\left(
\int_{-\infty}^A d\theta
{e^{-\theta} \over \cosh(\theta-\theta_B)^2}
\rho_+(\theta)
\right) .
}
This expression is more convenient that
the form-factors one of the previous section. In particular,
using the Fourier representation of $\rho$, $\chi_0$
can be written as a convergent series. For small voltage
this series is in $V/T_B$ (recall $T_B\propto \lambda^{1/1-g}$):
it corresponds to the approach of the IR fixed point
along the stress energy tensor. For large voltage,
the series is in $(T_B/V)^{1-g}$, and corresponds to the
(conformal)  perturbation of
the UV fixed point.
The leading term in that case is found to be
\eqn\larvtb{ \chi_0\simeq {2 (1-g)\over \sqrt{\pi} g^2 V}
 {1\over \Gamma(-g) \Gamma(g-1/2)
\cos\pi g}
\left( {T_B\over e^{A+\Delta}}\right)^{2(1-g)}.
}
By using the relation between $T_B$ and $\lambda$, this can be recast
as
\eqn\larvvtbb{\chi_0\simeq {2\Gamma(3-2g)\over \pi} \lambda^2
(gV)^{2g-3},}
in agreement with first order conformal perturbation theory.

The constant $T_B$ can be easily  eliminated to form universal
amplitudes. For example, the large and low voltage results
are related by the following universal product~:
\eqn\univg{\eqalign{
&\left[\lim_{V\rightarrow 0} \chi_0(V)\right]^{2(1-g)} \times
\left[\lim_{V\rightarrow\infty}
V^{3-2g} \chi_0(V)\right]=F(g) \cr
&F(g)={2\sqrt{\pi} (1-g)\over \pi^{5-4g} g^{4-2g}}
{1\over \Gamma (-g) \Gamma(g-1/2) \cos\pi g}
\left( {2 G_+(i)\over G_+(0) e^{\Delta}}\right)^{2(1-g)}.
}}

\subsec{Singularities}

When the impurity is present,  the finely
tuned cancellations
killing off the singularities will be upset by different R-matrices
terms. Let us
see how this works in the double well problem.

Consider first $\omega_c=gV$. Recall that in the absence of impurity,
there was a fine cancellation between
the particle-hole process below $\omega_c$ and the one-breather
process
above it. In the presence of the impurity, consider first
the particle-hole process. For the hole close
to
$\theta=-\infty$,
the reflexion matrix is simply equal to one. There is no dressing
effect because the
only particles in the sea affected by this hole are themselves at
rapidities
close to $-\infty$, where their $R$ matrix is also unity. Hence the
only part that sees the impurity is the particle close to the Fermi
rapidity.
For this one we have the dressed R-matrix \renrdef, so the
particle hole process contributes to $\chi''(\omega)$ by
a term  proportional to ~:
\eqn\singuyt{{\cal A}_{ph}\left\{R(A)\exp\left[\int_{-\infty}^A
d\theta
 F(\theta|A){d\over d\theta}\ln
R(\theta)\right]-1\right\}(\omega_c-\omega), \omega<\omega_c.}
At low voltage in particular this is proportional to  ${V\over T_B}
(\omega_c-\omega)$.

The breather reflection matrix for
$\theta_b=-\infty$ is
equal to unity, so  the breather process does not contribute linearly
to $\chi''(\omega)$ around $\omega_c$.
Hence \singuyt\ is actually the whole leading behaviour  close to
$\omega_c$, and there is now a singularity which we refer to as
being of the type $|\omega-\omega_c|$ (a discontinuity in the
first order derivative).

The same analysis can be carried out for the other processes.
However, because several processes  cross at $\omega_c=ngV, n> 2$,
and there are more amplitudes than relations to fix them, we cannot
be as conclusive: the singularity will be of the type $|\omega- ngV|$
unless some special cancellations occur, in which case it will be
weaker.

\newsec{The AC noise for tunneling in the fractional quantum Hall
effect}

\subsec{Low frequency behaviour}

The case of the FQHE is more complicated.  In the
dissipative quantum mechanics problem, the reflection matrices
 are antidiagonal
in the soliton/antisoliton basis.
For the impurity problem this is no longer true:  the reflection
matrix can connect
a state consisting of all right-moving
solitons to another in which there are left-moving solitons and
antisolitons.
Without a voltage, eigenstates obviously  read~:
\eqn\decomp{
||\theta>_\epsilon=|\theta>_\epsilon^R+R(\theta)_\epsilon^{\epsilon'}
|\theta>_{\epsilon'}^L.
}
At this stage, one must be very careful. We have argued
before that in the presence of a voltage, solitons and  antisolitons
have different energies, so it would seem that \decomp\ is
not an eigenstate of the hamiltonian when $V\neq 0$. This
conclusion is incorrect, for a subtle reason. The boundary
sine-Gordon model
must be   thought of as an anisotropic Kondo model
with a special (cyclic) representation of $SU(2)_q$ at
the boundary \ref\flskond{P. Fendley, F. Lesage, H. Saleur, J. Stat.
Phys.
(1996) to appear, cond-mat/9510055.}(or, more generally, a
representation of the ``q-oscillator algebra'' \ref\blz{V.V.
Bazhanov,
S.L.
Lukyanov, A.B. Zamolodchikov, Comm. Math. Phys. 177 (1996) 381,
hep-th/9412229.})
{}.
 The additional degrees of freedom
provided by this boundary spin
must be included in the proper definition of
asymptotic states, and in particular they reinstate spin (charge)
conservation. At the
end of the day, this auxiliary spin can be gauged away and
the result is that one can, indeed, treat \decomp\ as an eigenstate
with the
same energy as $|\theta>_\epsilon^R$. See the appendix for more
details.

More general eigenstates
are mixtures of left and right moving particles, which
we denote~:
$$
||\theta_1,...,\theta_n>_{\epsilon_1,...,\epsilon_n}
$$
In particular, the ground state gets modified, the sea of
solitons
becomes a sea made of superpositions of solitons and antisolitons
according to $|\theta>_+\to ||\theta>_+=|\theta>_+^R+
P(\theta)|\theta>_-^L
+Q(\theta)|\theta>_+^L$,
where we introduced the notation~:
\eqn\notar{
R^\epsilon_{\epsilon'}=\pmatrix{Q & P \cr P & Q},
}
and the elements $P,Q$ are given in \ref\FSW{P. Fendley, H. Saleur,
N. Warner, Nucl. Phys. B430 (1994) 577.}. When we compute the full
current-current
correlation function, different terms arise depending on the
chirality of the $J$ operators. They actually
have different physical meanings, and we will treat them separately.

To understand the physical meaning of the different terms, let us
stress again
that the  boundary formalism is fully equivalent to a formalism with
only R movers scattered through an impurity. The R state in \decomp\
can then be considered as an ``in'' R-state, and the L states in
\decomp\
as ``out'' R-states, with the same energy but possibly different
quantum numbers. Formula \decomp\ is thus
a scattering eigenstate in the traditional sense, and the
following computations are fully equivalent to Landauer-B\"uttiker
scattering \ref\LB{R. Landauer, Phys. Rev. B47 (1993) 16427; M.
B\"uttiker, Phys.
Rev. B46 (1992) 12485.} as carried out in \CFWI\ for the particular
case $g=1/2$.

Let us finally emphasize that we are only concerned here with
the $\omega$ dependent component of the noise: the DC noise (which
has been already computed in \flsnoise)  is implicitely subtracted.

The first term corresponds to $J_RJ_R$. This term sees
only the R-states in \decomp, and
can be thought of as a noise purely
in the imcoming channel: it  is  thus insensitive
to the boundary interaction. One has, at coincident points~:
\eqn\pompi{S_{RR}={g\over 2\pi}|\omega|.}

The second term corresponds to $J_LJ_R$ and is more complicated. It
can be thought of as a noise between incoming and outgoing channels.
$J_R$ acting on the
new ground state can create  pairs particle-hole of solitons,
create pairs
solitons antisolitons, or add breathers. Only the first process
contributes to low energy. The particle-hole pairs then ``bounce'' on
the boundary
where they can switch charges. They are then acted on by $J_L$ and
``destroyed''.
Observe however
that in this destruction, an initial soliton
can be replaced by an antisoliton since now the ground state
is a mixture.

The dressing follows a similar principle. If we write~:
$$
P^*(\tilde{\theta}_1)P(\theta_1)+Q^*(\tilde{\theta_1})
Q(\theta_1)\approx [1+ i\phi(\theta_1)(
\tilde{\theta}_1-\theta_1)],
$$
(where $\phi$ is real as can easily be proven using
that $|P|^2+|Q|^2=1$) we can dress the reflection matrices according
to~:
\eqn\bizz{\eqalign{{\cal Q}(\theta)&=Q(\theta)\exp i
\left[\int_{-\infty}^A
F(\theta'|\theta)\phi(\theta')d\theta'\right]\cr
{\cal P}(\theta)&=P(\theta)\exp i\left[\int_{-\infty}^A
F(\theta'|\theta)\phi(\theta')
d\theta'\right].\cr}}

{}From \LSS , we find that the low frequency noise is now written~:
\eqn\realgood{\eqalign{S_{LR}(\omega)(x_1,x_2)=&\int_{-\infty}^\infty
dt e^{i\omega t}\left< J_L(t,x_1)J_R(0,x_2)\right>\cr
=&{g\over 2\pi}\int_{-\infty}^{\ln\omega}d\beta_2d\beta_2'
\left[{\cal P}^*(\theta_2){\cal P}(\theta'_2)-{\cal
Q}^*(\theta_2){\cal Q}(\theta'_2)\right]
e^{\beta_2}e^{\beta_2'}\times \cr & \ \
\delta\left(\omega-e^{\beta_2}-
e^{\beta_2'}\right) e^{i\omega(x_1-x_2)}.\cr
}}
here $x_1$ and $x_2$ physically correspond to
the two sides of the impurity, and should be taken different
in general. For the tunneling noise, both $x_1$ and $x_2$ will
approach $0$ and the phase will disappear.  By the same manipulations
as above this can be rewritten~:
\eqn\realgoodi{S_{LR}(\omega)(x_1,x_2)=
 {g\over 2\pi}e^{i\omega(x_1-x_2)}\int_0^\omega \left[{\cal
P}(x){\cal P}^*(x-\omega)-
{\cal Q}(x){\cal Q}^*(x-\omega)\right]dx.}
Of particular interest is the limiting behaviour of $S$ as $\omega\to
0$. In that limit,
we pick up the value of the argument in the integral \realgoodi\ for
$x=0$ for which
the dressing effect just cancels out. One finds simply~:
\eqn\mainu{\eqalign{S_{LR}(\omega)&\approx {g|\omega|\over
2\pi}\left[|P(A)|^2-|Q(A)|^2\right]e^{i\omega(x_1-x_2)}\cr
&={g|\omega|\over 2\pi}
(2|P(A)|^2-1)e^{i\omega(x_1-x_2)}\cr
&={g|\omega|\over\pi}\left({e^{2({1\over g}-1)
A}\over e^{2({1\over g}-1)A}+T_B^{2({1\over g}-1)}}-{1\over
2}\right)e^{i\omega(x_1-x_2)}.\cr
}}
At large voltage it goes to the noise without impurity as expected.
At small voltage we find~:
\eqn\smallvmain{S_{LR}(\omega)\approx
e^{i\omega(x_1-x_2)}\left[{g|\omega|\over\pi}\left({G_+(0)\over
2G_+(i)}\right)^{2({1\over g}-1)}
\left({V\over T_B}\right)^{2({1\over g}-1)} - {g|\omega|\over
2\pi}\right].
}

The last term corresponds to $J_LJ_L$ and can be interpreted
as a noise purely in the outgoing channel. It does have quite a bit
of structure. At leading order
at low frequencies, the first process that
contributes is  the creation of a particle hole pair near
the Fermi surface. Observe however that due to the mixing
induced by the boundary, one can either destroy
a soliton and create another one above the sea,
or destroy an antisoliton and create another one above the sea
\foot{Due to this mixing also, it is also  possible  to destroy a
{\bf pair}
of particles close to the Fermi surface. This process
however does not contribute at leading order because the form factors
for destroying a pair $+-$ and a pair $-+$ are opposite.}.   At
leading
order, one finds, at coincident points~:
\eqn\pompii{S_{LL}(\omega)\approx {g|\omega|\over 2\pi} \left(
|P(A)|^2-|Q(A)|^2\right)^2={g|\omega|\over 2\pi}
\left(1-{4\over\pi}|P(A)|^2|Q(A)|^2\right).}
Observe that $S_{RR},S_{LR}$ and $S_{LL}$ are respectively of order
$0,2$ and $4$ in the R-matrix elements. Adding all the components we
find the
low frequency noise of the tunneling current~:
\eqn\pompiii{S_{t}(\omega)\approx {g|\omega|\over 2\pi} |Q(A)|^4=
{g|\omega|\over 2\pi}\left(
{T_B^{2({1\over g}-1)}\over e^{2({1\over g}-1)A}+T_B^{2({1\over
g}-1)}}\right)^2,\  \omega\to 0.
}
This noise reproduces the standard result $S_t={g|\omega|\over 2\pi}$
at small voltage, where there is no transmitted
current. This noise vanishes at very large voltage, when the
tunneling current goes to zero, as~:
\eqn\limnois{S_t(\omega)\approx {g\over 2\pi}\vert \omega \vert
\left(
{2G_+(i)\over G_+(0)}\right)^{4({1\over g}-1)}
\left({T_B\over V}\right)^{4({1\over g}-1)},\omega\ \to 0.
}
The noise for the current running down the sample
can also be expressed simply as~:
\eqn\otherlimnois{S_{T}(\omega)\approx {g|\omega|\over 2\pi}
|P(A)|^4.}

Let us stress here  that  a priori, all these results hold only
for $0<g<1$. As $g\to 1$, the
threshold for
adding an antisoliton vanishes $\hbox{ Min
}\left(\epsilon_-\right)=(1-g)V\to 0$, so exactly at $g=1$
other processes are implied in the low frequency physics. In other
words, we expect a priori that
\eqn\surpr{\lim_{\omega\to 0} \lim_{g\to 1}
{S(\omega)\over|\omega|}\neq
 \lim_{g\to 1}\lim_{\omega\to 0} {S(\omega)\over|\omega|}.}
Interestingly however, exactly for $g=1$ the noise
is easy to compute directly \ref\EY{S. R. E. Yang. Solid State Comm.
81 (1992) 375.}
since the electrons are non interacting, and one finds the same
as \limnois. In that light, \pompiii\ actually
appears very natural since we showed that the low energy
excitations at $V\neq 0$ are described by an effective
free fermion theory.

Similarly, as $g\to 0$, the threshold for adding breathers vanishes,
and non commutativity of the limits is also a priori expected.

\subsec{Other singularities?}

The effect of thresholds we have
discussed for the double well problem should be
observable as well in the noise, both for $S_{LR}$ and $S_{LL}$,
leading to a first  singularity of the form $|\omega-gV|$, and
presumably
other singularities of the same formm $|\omega-ngV|$.
As an example, let us discuss the singularity at
$\omega=gV$. Like in the double well
problem, it arises because  the two processes of creation of a pair
particle-hole and of a breather do not match at $\omega=\omega_c$.
For the particle-hole process,
the bulk amplitude gets multiplied by a factor $|{\cal Q}(A)|^2$.
This is
because, for a hole at $\theta_1$ and a particle at $\theta_2$,
the general amplitude is~:
$$
\left|{\cal P}^*(\theta_2){\cal P}(\theta_1)-
{\cal Q}^*(\theta_2){\cal Q}(\theta_1)\right|^2,
$$
and at leading order close to the threshold,
 we have $\theta_2=-\infty$ and $\theta_1=A$. For the breather
process, the bulk
 amplitude is unchanged since at leading order, the breather
reflection
matrix appears only in the form of its modulus square. Hence the
 singularity is proportional to
\eqn\slop{\left[|1-{\cal Q}(A)|^2\right]|\omega-gV|.}

In addition, other singularities take place
in $S_{LR}$. This is because, beside the processes
of the double well problem, other  processes are allowed here,
for instance the destruction of a
{\bf pair} of particles close to the Fermi surface,
as already mentioned earlier.  Now this process
can take place provided the frequency $\omega$ is smaller than
twice the maximum energy of a hole in the sea, or $\omega<2gV$.
Hence,
this process  contributes a singularity which is
however weaker
than $|\omega-2gV|$. More generally, processes
where $2p$ particles
are destroyed in the sea have a maximum
frequency $\omega=2pgV$, and should lead to
singularities weaker than $|\omega-2pgV|$.
Recall
that such processes are allowed in the general interacting theory, in
contrast
with the case $g=1/2$ where only pairs can be created by the current,
which implies that  the excess noise in the outgoing channel vanishes
for $\omega>2gV$.

\subsec{Comparison with  perturbative results}

Finally, we would like to compare our results with the
perturbative approach of \CFW, \CFWI\ concerning the noise in the
tunneling problem.  For the low frequency behaviour,
we agree that there is a singularity of the type $|\omega|$, but
the amplitude \limnois\ that we find does not
agree with these authors, except for $g=1/2$ and $g=1$. Observe that
the amplitude we find for the $|\omega|$ singularity, while
expanding nicely at large $\lambda$ in powers of $\lambda^{4/g}$,
does
not expand in powers of $\lambda^4$ at small $\lambda$. This suggests
that the UV perturbation theory attempted in \CFW,\CFWI\ has
a vanishing radius of convergence. In particular,
we find that the large voltage behaviour of the amplitude
goes as $V^{4-4/g}$, instead of $V^{4(g-1)}$.

In fact, we can discuss the difference in more details. Indeed,
 Chamon, Freed and Wen have recently \ref\CFWII{C. de C. Chamon, D.
Freed,
and X. G. Wen, to appear} obtained
for the noise the expression \foot{Unfortunately, what is called L
and R here
is a bit different from the notations adopted in \CFWI,\CFWII. The
different
noises are however simply related by linear  combinations.}
\eqn\fcw{S_T(\omega)\approx {\omega\over 2g\pi} \left({dI\over
dV}\right)^2,}
instead of our expression \otherlimnois. It is worthwhile
to examine the difference in more details.  In \flsbig\ the DC
current in the presence of a voltage was determined,
\eqn\alli{I=2\pi\int_{-\infty}^A \rho_+(\theta)|P(\theta)|^2.}
To obtain the differential conductance, we take derivative with
respect to $V$.
The derivative of the density can be obtained from \denseq:
\eqn\allii{2\pi{d\rho_+(\theta)\over dV}=\sqrt{{g\over
2}}L(\theta,A),}
from which it follows that
\eqn\diffcond{G={dI\over dV}=\sqrt{{g\over
2}}\left[|P(A)|^2+\int_{-\infty}^A
L(\theta,A)|P(\theta)|^2d\theta\right].}
Define a new renormalized reflection matrix
\eqn\newrens{|{\cal
P}_{diff}(\theta)|^2=|P(\theta)|^2-\int_{-\infty}^A
\Phi(\theta-\theta')|{\cal P}_{diff}(\theta')|^2d\theta',}
then \diffcond\ reads
\eqn\newdiffcond{G=\sqrt{{g\over 2}}|{\cal P}_{diff}(A)|^2.}
The meaning of this result is as follows. Starting with a  voltage
$V$,
one is interested in the additional current when $V\to V+\delta V$.
This current
has two origins: one is the shift of the Fermi sea, the other one is
the
change in populations deep in the Fermi sea. The shift of the Fermi
sea
adds a number of solitons
\eqn\addnum{\delta N=L\rho_+(A)\delta A={L\over 2\pi}\sqrt{{g\over
2}}\delta V.}
If we want to write the change of current in terms of those particles
only, it has to involve a dressed reflection matrix
which takes into account all what happens deep in the Fermi sea:
\eqn\expdiffcond{{\delta I\over \delta V}=2\pi {\delta N\over L}
|{\cal P}_{diff}(A)|^2,}
which coincides with \diffcond\ using \newrens. Hence the
differential conductance
can be fully interpreted in terms of  a  new dressed reflection
matrix.

In this language, the formula  \fcw\ reads
\eqn\newfcw{S_T(\omega)\approx {g\omega\over 2\pi}|{\cal
P}_{diff}(A)|^4.}

The  difference with  our formula \otherlimnois\   is thus fully a
difference
 in the dressed reflection matrix. Naively, based on current
computations,
one would have expected (eg by analogy with dissipative quantum
mechanics) that
 \newfcw\ would hold. The key  point however is that the  dressing of
reflection matrices is not a universal property, but depends
on the quantity under study. For the conductance, the whole of the
Fermi sea matters, and the renormalized reflection matrix   ${\cal
P}_{diff}$
involves the whole sea as well. For the ($T=0$) noise,
the only effect of the sea in ${\cal P}$ is a phase
that disappears in moduli square.
Presumably, this is a non perturbative effect that cannot be seen in
the approach
of \CFWII.

Similarly, we
disagree on the singularity structure. Except for $g=1/2$ we find
that the noise $S_t(\omega)$  has a singularity at $\omega=gV$ -
 the ``quasi particle singularity'' -
where the first derivative is discontinuous,
while the
authors of \CFW, \CFWI\ argued that
this singularity was either a $g$-dependent power law, or was
absent.

\newsec{Conclusions.}

This paper
is a first step towards the computation
of correlation functions in quantum impurity problems
in the presence of a voltage and a temperature.
While much remains to be done, we think the present results
already indicate very interesting features, and could lead to
numerical
and experimental applications. In particular, the presence of
a singularity at $\omega=gV$ at $T=0$,
while probably  unobservable experimentally, should lead
to a pic in the derivative ${dS(\omega)\over d\omega}$ at finite $T$.
 We also expect
that such pics should appear in fact  at  regularly
spaced values $\omega=ngV$ at finite $T$ (while presumably
the amplitudes of these pics will decrease very rapidely with $n$).
The meaning of the singularities at $T=0$ can be understood in
the double well problem:
$\chi''(\omega)$ going as $|\omega-gV|=|\omega-\epsilon|$
simply means that the long time behaviour of the spin spin correlator
has an oscillatory  component ${e^{i\epsilon t}\over t^2}$.

A more complete discussion of the low energy
excitations,  including further justifications of \renslim,
will be provided in the following paper \LSnext, together with
some results at $V=0$ but $T>0$.

\vskip 0.4cm
\centerline{\bf Acknowledgements}

We thank C. de C. Chamon, S. Chakravarty, T.A. Costi,  D. Freed
and A. Rosch for very
useful discussions, and for
kindly communicating their results prior publication. We especially
thank C. de C. Chamon for patiently
explaining to us the papers \CFW,\CFWI,\CFWII, and T. A. Costi and
A. Rosch for checking some of our formulas at high field using
their NRG method.

This work was supported by the Packard Foundation, the
National Young Investigator program (NSF-PHY-9357207) and
the DOE (DE-FG03-84ER40168).
F. Lesage is also  partly
supported by a canadian NSERC Postdoctoral Fellowship.

\appendix{A}{The effects of the magnetic field and voltage}

Introduce~:
\eqn\kondi{L_G={1\over 16\pi g}\int_{-\infty}^0 dx \
\left[(\partial_t \phi)^2
-(\partial_x \phi)^2\right].}
Let us then define
\eqn\newdefapp{L(h')=L_G-{h'\over 4\pi}\int_{-\infty}^0
\partial_t\phi .}
Introducing
\eqn\hamkondi{H_G={1\over 2}\int_{-\infty}^0 \left[8\pi g\Pi^2
+{1\over 8\pi g}(\partial_x\phi)^2\right],}
the associated hamiltonian is
\eqn\newdefham{H(h')=H_G+L{g\over 4\pi}(h')^2+2gh'\int_{-\infty}^0
\Pi(x)dx.}
More generally, we can also put a boundary coupling and a boundary
field,
defining
\eqn\bdrham{H_B=\lambda\left(S_+e^{-i\phi(0)/2}+S_-e^{i\phi(0)/2}
\right),}
and
\eqn\hamdefff{H(h,h')=H_G+H_B+L{g\over
4\pi}(h')^2+2gh'\int_{-\infty}^0 \Pi(x)dx+{h\over 2}S_z,}
The unconventional normalisation is taken to agree with the
presentation in the bulk of the text. The total z-component of the
spin, which
commutes with the hamiltonian, reads with these normalizations
\eqn\totalspin{S_{total}={1\over 2\pi
g}\int_{-\infty}^0\partial_t\phi+S_z.}
Hence, when $h'={h\over g}$, \hamdefff\ describes the field ${h\over
2}$ coupled to the total spin. Otherwise, a different field is
coupled to
the boundary and bulk components of this spin. Now, the term linear
in $\Pi$
in \hamdefff\ can be eliminated by a unitary transformation.
Introduce
\eqn\defujnit{U=\exp\left[{ih'\over 4\pi}\int_{-\infty}^0
\phi(x)dx\right],}
then one finds
\eqn\main{U H(h,h') U^{-1}=H(h,h'=0)-L{g\over 4\pi}(h')^2.}
As a result, the thermodynamic
properties
of the impurity are  the same whether there is a field coupled to
the
impurity spin $S_z$ only, or a field coupled to the impurity
and  another field coupled to the the component
$\int_{-\infty}^0\partial_t\phi$, including the
particular case of a field coupled to the total spin.
This is similar to the
observations in
\ref\Low{J. H. Lowenstein, Phys. Rev. B29 (1984) 4120.}.
To be more precise, introduce
\eqn\defikon{{\cal Z}_1(h,h')=Tr e^{-\beta H(h,h')},}
with the trace taken in the spin $1/2$  representation (where by
convention recall we chose $S_z=\pm 1$). Then one has
\eqn\rela{Z_{1}(h)={{\cal Z}_{1}(h,h',\lambda)\over {\cal
Z}_{1}(h=0,h',\lambda=0)}=
{{\cal Z}_{1}(h,h'=0,\lambda)\over {\cal
Z}_{1}(h=0,h'=0,\lambda=0)}. }
In \flskond, we checked this relation
in the particular case $h'={h\over g}$ in the first numerator, since
we found
in \flskond\ that the coefficients of the perturbative
expansion in $\lambda$ of the second ratio in \rela\  coincide
exactly with
the results of TBA calculations for the Kondo model coupled to the
{\bf total} spin, ie the first ratio in \rela.

In \totalspin, the $g$ factor is
a standard renormalization that can be observed for instance in the
continuum
limit of the XXZ chain regularization  \KR\ of the Kondo model
\ref\hs{H.
Saleur,
unpublished.}. For convenience, we define the spin of a
soliton or antisoliton to be $\pm {1\over g}$ in what follows.

The scattering theory description of the spin $1/2$ Kondo model
involves only the soliton and antisoliton because
it is an IR description, and in the IR the boundary spin is screened.
Conservation of the spin then simply translates into the fact
that solitons bounce back into antisolitons.

A coupling ${h\over 2}S_z$ to the boundary spin can also be traded
for
a time dependent boundary coupling. Indeed, consider
for instance the computation of ${\cal Z}_1$ perturbatively in
$\lambda$.
At order
$2n$, one has Coulomb gas integrals related with a 2D Coulomb gas on
a circle
with $n$ positive and $n$ negative charges alternating. Call
$\tau=it$
the imaginary time, and suppose an $S_+$ term has been
inserted at $\tau$ and an $S_-$ term at $\tau'>\tau$.  The effect of
the
term ${h\over 2}S_z$ in the hamiltonian
\hamdefff\ is to give a weight $\exp -(\tau'-\tau)h$
to the pair of insertions
of $S_+,S_-$, relative to the
weight with no insertion. Now from \kondi\ we see that every
insertion of
$S_\pm$
is coupled with an insertion of $\exp\mp i\phi(0)/2$. The effect
of
the  ${h\over 2} S_z$ term can thus be absorbed
into a time-dependent phase  multiplying the vertex
operators in $H_B$,
\eqn\timedeppha{S_{\pm}\to S_{\pm}e^{\pm ith}.}
In addition, the term
 ${h\over 2}S_z$
gives rise to an overall  term $\exp \pm \beta h/2$, independent
of the order $n$,  that can be absorbed by taking a modified trace
$Tr (.)\to Tr (. )\ e^{{h\over 2}S_z}$.

The foregoing chain of arguments  generalizes
to higher spins in $SU(2)_q$. This is obvious for the $h'$ term
which is independent of the spin. As
for the $h$ term,
it follows simply from the fact that insertions of $S_\pm$ change
the spin $S_z$ by $\pm 1$,
independently of the representation.

The properties of correlators with \hamdefff\ can be analyzed by
the same sort of arguments, and the various transformations
easily followed on each term of the correlators.

The first relevant point for
this paper is that, although the dissipative quantum mechanics
(anisotropic Kondo) hamiltonian  has a field coupled  only  to the
boundary spin
 $h=\epsilon\equiv gV$, it
can be transformed, up to a shift of the ground state energy, into  a
hamiltonian where the field is coupled
to the total spin $h'={\epsilon\over g}=V,\ h=\epsilon=gV$, and  then
easily
diagonalized  by Bethe-ansatz.  In the massless description
we are using, the boundary spin is  screened, and only solitons and
antisolitons have to be considered, with an
energy shift equal to $\pm V/2=\pm \epsilon/2g$. Since we are using
a scattering approach, the eigenstates are
built as explained
at the beginning of section 4 by combining the left and right
(asymptotic)  eigenstates
of the full line hamiltonian \hbulk. The presence of
the impurity  does actually change the
densities by a factor of order $1/L$, which does not affect
the results we are interested in.

The second point  concerns the impurity in a Luttinger
liquid. Recall, following \flskond, that the boundary sine-Gordon
model is in fact equivalent
to an anisotropic Kondo model where the boundary spin
is defined in a special, cyclic  representation of $SU(2)_q$ (or,
alternatively, in a representation of the q-oscillator
algebra \ref\lykia{V.V. Bazhanov,
S.L. Lukyanov, A.B. Zamolodchikov, hep-th/9604044.}).
While these boundary spins
can be ``gauged away'', let us discuss them a bit more. First,
asymptotic states are now
characterized by particles {\bf and} by the value of the renormalized
boundary spin.  Indeed, let us stress here that the boundary spins
that appear in
the hamiltonian $s_H$ and in the  scattering theory $s_{scat}$ differ
for two reasons. There is  the  screening, already mentioned
previously for the
Kondo problem, and the multiplicative renormalization, which can be
observed by considering the energies with a magnetic field. One finds
$s_{scat}={1\over g}(s_H-1)$, where the $g$ factor was also mentioned
previously.

The main use of the boundary spins is to reinstall conservation
of the  spin. This is obvious on the hamiltonian
which looks then like a higher spin Kondo problem. This
is also obvious in the scattering theory : every time a soliton
bounces back as a soliton instead of an antisoliton, the boundary
spin increases its value
accordingly.
The conservation of the spin allows
diagonalization of the problem with a voltage. In particular since a
pair R-soliton,
 $S_z={m\over g}$ and L-soliton, $S_z={m+1\over g}$
have the same  charge and the same kinetic energy, they should
have the same energy, ie $\epsilon_--{V\over 2}=\epsilon_++{V\over
2}$, where we used $h=gV$, an identity
which we mentioned previously. Also, in section 4, while the state
\decomp\ seemed
to be made of  particles reflected  with different energies,
it is more precisely  a combination of true asymptotic states that
involve also the boundary spin, that all have the same total spin,
and all the same energy.
Therefore, for the impurity problem, one starts with
a time dependent
term in the boundary coupling  in \hallbound. This term is then
transformed into
a field  $h=gV$ applied to the boundary spin. An additional term
coupled
to the component $\int_{-\infty}^0\partial_t\phi$ is then added, and
allows for the use
of the Bethe ansatz with a field coupled to the total spin. The
boundary spin is finally gauged away.

The arguments of this appendix provide the missing steps
in the proof of the DC noise formula (27) of \flnoise. They also
justify why in \flskond\ the Bethe ansatz for a field
coupled to the total spin was used, although the initial
hamiltonian had a field coupled to the boundary spin only.

Finally, let us stress that, while the double well
is a problem in equilibrium, the tunneling problem is not. As simple
way to see that is to observe that,
while we used an eigenstate of the boundary hamiltonian with the R
part made only of solitons, there are, thanks to the boundary
spin which can absorb all
spin (charge) shifts, an infinity of other boundary states, and
averaging over all of them would, for instance, lead to a vanishing
DC current
as expected in equilibrium. Our approach  here is fact fully
equivalent to the Landauer-B\"uttiker scattering.
In that context, it is also interesting to mention that the
transformations  previously discussed change the boundary
conditions at infinity, which can be interpreted  in terms of
reservoirs. The latter are implicitly there in the scattering
approach \CFWI.

\listrefs

\bye